# Machine Learning in Congestion Control: A Survey on Selected Algorithms and a New Roadmap to their Implementation

Zhilbert Tafa[1] and Veljko Milutinovic[2]

[1]Department of Computer Science, University for Business and Technology,
Prishtina, Kosovo
[2]Department of Computer Science, University of Indiana,
Bloomington, IN, USA

## Abstract

With the emergence of new technologies, computer networks are becoming more structurally complex, diverse and heterogenous. The increasing discrepancy (among the interconnected networks) in data rates, delays, packet loss, and transmission scenarios, influence significantly the dynamics of congestion control (CC) parametrization. In contrast to the traditional end-to-end CC algorithms that rely on strict rules, new approaches aim to involve machine learning in order to continuously adapt the CC to real-time network requirements. However, due to the high computational complexity and memory consumption, the feasibility of these schemes may still be questioned. This paper surveys selected machine-learning based approaches to CC and proposes a roadmap to their implementation in computer systems, by using dataflow computing and Gallium Arsenide (GaAs) chips.

## 1 Introduction

Congestion Control is a fundamental issue in computer networks. The objectives of CC can be presented in the three-dimensional space bounded by: resource utilization, performance, and fairness. The parametrization of algorithm inputs and continuous adjustment of control variables to meet the optimal point in this space is not an easy task.

Traditional end-to-end CC algorithms that have been implemented so far rely on predefined rules. Based on these rules, congestion window variable (cwnd) and Retransmission Time Out timer (RTO) are recalculated continuously. Depending on the congestion indicator, they can be classified into *loss-based*, *delay-based*, and *hybrid*. Many algorithms that belong to one of these types have already been developed and implemented. However, modern networks are complex, diverse and heterogenous in their structure. Applying strict rules in a multiparametric and highly dynamic environment does not perform equally well in different network conditions; i.e., the rules that perform near-optimal in one environment may not perform well in other environments or when network conditions change rapidly.

With regards to the problem space, as well as the diversity and variability of input parameters, new learning-based approaches have recently attracted a great research attention. Different from traditional CC algorithms, learning-based schemes are based on real-time network states to make decisions instead of using predetermined rules [1].

This paper is a short and concise survey of the representative learning-based congestion control algorithms, their main strengths and weaknesses as well as the opportunities of their real-time implementation.

In section 2, the essence of the (CC) problem is shortly presented.

In section 3, each of the selected algorithms are uniformly described by relying partially on the methodology given in [2]. More precisely, the description is focused on the following issues:
1. What are the specifics of the algorithm.
2. Why is it better as compared to others from literature.
3. To what extent is it better and under which conditions.
4. What has been done so far to prove its superiority.

Sections 4 present dataflow and GaAs paradigms in light of their capability to solve key computation obstacles of learning-based CC algorithms' implementation, respectively. Finally, section 5 concludes the paper.

## 2   Problem statement - issues with the existing CC

The essence of network congestion problem may be summarized as follows. The output links of network nodes (e.g., routers) are of limited bandwidth. A number of flows may enter a specific node and may need to be sent throughout a single interface. This discrepancy in arrival rate and output capacity is amortized by buffers, and may lead to the formation of queues. The packet buffering leads to increased packet delivery delays. If the participating hosts would continue sending packets at the same or higher data rates, the delays would extend to infinity and packet loss would occur.

To minimize or avoid this effect, CC algorithms continuously adjust their transmission parameters (such as cwnd or sending rate) as a function of network congestion indicators. An ideal congestion control algorithm will fully utilize the bottleneck bandwidth, yet keeping queuing delays to the minimum by not overfilling the "pipe" [3]. Precisely, the adjustments aim at keeping the network transmission at the optimal point bounded by: a) best performances (e.g., high throughput, low latency, and low packet loss rate), b) minimum or no congestion, and c) fair bandwidth sharing with other streams.

In the end-to-end CC approach, hosts can only implicitly sense the congestion, by relying on feedbacks from the network (such as the number of ACKs, their timing, packet delivery rate, etc.). With regards to the way the congestion is recognized at the sender, traditional CC algorithms are divided into three groups: loss-based, delay-based, and hybrid.

**Loss-based** CC algorithms take actions based on received (or unreceived) acknowledgments (ACKs), as indicator of packet loss. The congestion is considered severe if the RTO has expired before the ACK has arrived. Another congestion indicator as used by these algorithms is the reception of three duplicate ACKs. The aforementioned two situations will trigger different behavior of an algorithm. Loss-based algorithms are generally characterized as aggressive (as they probe for the network resources more aggressively). In their structure, they lack in two main things: a) since they wait for RTO or DupACKs, they detect congestion pretty late, after it gives even more negative effects on the network performances (by dropping packets) or after waiting for RTOs to expire, b) they cannot distinguish between packet loss due to congestion and packet loss due to, for instance, wireless fading, which makes these algorithms unsuitable for wireless communications (IoT, Wireless Ad Hoc Networks, etc). Various loss-based CC algorithms have been proposed and implemented for decades, such as Reno [4], Cubic [5] etc.

**Delay-based** CC algorithms rely on continuous Round-Trip Time (RTT) measurements to adjust data rates. If RTT begins to increase, a delay-based CC algorithm will reduce its sending rate. As such, these algorithms are able to react before congestion worsens, i.e., at the moment the queuing starts to increase and before packet loss occurs. Consequently, delay-based CC algorithms will cause much fewer packet transmissions than those which belong to the loss-based group. However, there are two main drawbacks related to the way delay-based CC algorithms work: a) they rely on RTT, but RTT estimation can be biased by routing dynamics, delayed ACKs, etc., and therefore may not be precise, b) since they are less aggressive in probing network resources, they will lack in bandwidth utilization as compared to the loss-based algorithms, especially in high-speed networks where transmission speed needs to be increased as fast as possible. Various delay-based CC algorithms have been proposed, such as Vegas [6] and its variants, but they have not been widely deployed since they cannot survive with loss-based algorithms.

Finally, **hybrid algorithms** aim to balance between the two approaches by increasing the transmission rate as quicky as possible when the network is underutilized (e.g., in high bandwidth networks), and by becoming more conservative when network utilization approaches to its limits. The general strategy of these algorithms is to estimate the delivery rate over time. Among many algorithms of this type that have been developed so far, Bottleneck Bandwidth RTT (BBR) [7] has shown best performances. It has been implemented or made available in entities such as Dropbox, Linux, and Google. The main problem with the hybrid protocols is that they are slow in achieving bandwidth utilization point. This property makes them unsuitable for short flows. Also, hybrid algorithms are often shown to be unfair when used together with other algorithms. Some of the observations on BBR are given in [8].

In addition to the aforementioned (structural and implementation-related) issues of the traditional CC algorithms, some studies [9, 10] show that TCP performs bad at adapting to dynamic network conditions, such as fast bandwidth variations. In [11], cognitive radio spectrum sensing and switching turns out to lower the TCP NewReno bandwidth utilization drastically. Additionally, simulations given in [12] lead to some more general observations and limitations regarded to all three CC categories, such as: a) none of the representative algorithms are able to fully utilize the available bandwidth, even when no additional traffic is present on the link, b) delay-based and hybrid algorithms will show significant performance deterioration when used together with flows from loss-based group of algorithms, c) all conventional algorithms are sensitive to different RTTs of the flows. The last limitation represents a situation when different flows have different RTTs. The one with smaller RTT will probe for resources more often. Consequently, it will have its cwnd or data rate growing more aggressively at the same time suppressing sending rates of the flows which are based on other algorithms.

We can now identify a common problem with the traditional CC algorithms: they map out a diverse (continuous) dynamic input space received from a wide range of networking technologies into a strict output space of a very limited number of control parameters and rules. As such, various traditional solutions can be considered as near-optimal only for a small number of scenarios, in networks with low dynamics. The presented limitations may be minimized or eliminated if better adaptability to the network dynamics would be provided. These premises potentially locate the ML-based approaches to the heart of CC.

## 3 Selected machine learning algorithms for CC

Machine learning has recently attracted great attention in addressing the traditional rate-adjusting CC issues. Because of their ability to recognize the nature of data and to find outputs that fit best to some predefined goals, under various networking environments and conditions, they show potential to improve both steps of CC, namely: a) recognizing the nature of the congestion, b) adjusting the actual sending rate/cwnd to the most optimal value.

There are three main ML techniques: supervised learning (SL), unsupervised learning (UL), and reinforcement learning (RL).

In CC context, we deduce that ML algorithms can be used in two ways: a) as an attachment/improvement to the existing CC algorithms, b) as new algorithms that adapt CC actions without relying on pre-defined rules. In this mini-survey, we discuss the deployment of the representative algorithms from both groups, based on their efficiency and their up-to-date contribution to the CC. In the selection process, the research trends and the frequency of usage of each ML technique in solving aforementioned CC issues were also taken into account. Accordingly, more weight and space are given to the RL, SL is more narrowly presented, while UL is omitted from the discussion.

### 3.1 Improvement of the existing CC schemes by using SL algorithms

Supervised learning algorithms build their models based on previous experience. They are suitable for classification and prediction when some previous data inputs and outputs are known. This group of algorithms iteratively learn to adjust their internal parameters in order to build a model that should be able to successfully classify future (new) data or predict their outputs. As such, supervised learning can be useful in classification and prediction of specific CC issues. E.g., loss type classification and delay prediction are of crucial importance to the efficiency of CC algorithms. Traditional CC algorithms are not able to make a distinction between congestion induced loss and loss due to the erroneous link. Consequently, they will decrease the sending rate even if the network is not actually congested, which will lead to the inefficient use of bandwidth and network performance deterioration.

Learning-based classification group of algorithms may be a good candidate to help existing CC algorithms solve some classification and prediction issues. They can help the widely implemented protocols to distinguish real congestion from false alarms and to react accordingly. Based on the specific network behavior (e.g., one way packet delay, inter-packet time, etc.), the algorithms can decide whether a specific condition is induced by the congestion or by some other factors. If packet loss is caused by congestion, a system can use ordinary rule-based actions (e.g., halving the cwnd etc.). If a packet loss is caused by some other sources, then no action is taken. This patch to the existing CC has a potential to significantly improve bandwidth utilization. However, misclassification of a congestion-induced packet loss may lead to no reaction of the CC, in which case congestion can get even worse. To improve accuracy, joint implementation of different ML algorithms can be used [13].

Another domain where SL algorithms may contribute to the improvement of widely implemented algorithms is delay prediction. As mentioned, some CC algorithms use delay as a congestion signal; where RTT prediction is of crucial importance. Traditional RTT estimation is based on the Exponential Weighted Moving Average (EWMA) technique. However, this metrics is not appropriate in highly dynamic networks. More efficient prediction of an upcoming delay will significantly improve the algorithm's reaction time. E.g., based on the relationship between RTT and sending rate, supervised learning algorithms can more accurately predict the delay and hence improve CC algorithm to timely detect the congestion.

Traffic type and network bandwidth are also of great importance to the efficiency of the traditional TCP CC, while traffic type is highly correlated to the flow length. E.g., web browsing is characterized by short flow transmissions, while large file downloads are characterized by long flows. It has been shown that, most of the time, traffic flows are of short duration. On the other hand, traditional CC starts with a small cwnd value, which disables the short flows to utilize the bandwidth during their lifetime. Instead of sending several segments per RTT, the sender may have to wait many RTTs for the cwnd to increase. Some TCP implementations propose to start with higher initial window, but it cannot be appropriate in low-speed networks. In this context, we consider SL as having a good potential to improve the existing TCP implementations. For example, SL can make the appropriate classification on flow types and network characteristics, and can initiate cwnd accordingly.

Finally, SL algorithms can help hosts decide which of the existing algorithms is more appropriate to be used given the specific flow and specific network conditions. Precisely, various TCP flavors have shown to be more efficient for different flow types and in different network environments (delay, bandwidth, etc.). For instance, it has been shown that BBR will generally perform better than other protocols in networks of high bandwidth and low delays, while Cubic is more appropriate for long flows over WAN. Also, over cellular networks, C2TCP (Cellular Controlled Delay TCP) and Westwood would overperform other algorithms for long flows and short flows, respectively. SL algorithms can be trained to detect the network conditions and to choose a specific algorithm for a specific flow. One such approach is given in [14].

To summarize on the above discussion, by identifying and predicting network congestion indicators and traffic type more accurately, or by dynamically selecting the most efficient CC algorithm under a given networking conditions, SL algorithms can present an important patch to the existing CC implementations. Finally, various SL algorithms can be jointly used to upgrade the existing CC approaches to a higher level of efficiency.

### 3.1.1 Decision Trees

Decision Trees (DT) are of the oldest but still of the most popular SL algorithms. Decision process goes through branching, from the most significant attribute to the least significant one. Classification and Regression Trees (CART) [15] measure the "goodness" $\Phi(s|t)$ of candidate split attribute *s* at node *t*, where:

$$\Phi(s|t) = 2P_L P_R \sum_{j=1}^{\#classes} |P(j|t_L) - P(j|t_R)|$$

where, $t_L$ is left child node at *t*, $t_R$ is right child node at *t*, $P_L$ and $P_T$ measure the number of records at $t_L$ and $t_R$ per overall number of records in the training set, respectively. Finally, $P(j|t_L)$ and $P(j|t_R)$, are the measures of number of class j records at $t_L$ and $t_R$ per number of records at t, respectively.

Another DT variant - C4.5 and its inheritor C5.0 rely on information gain. The entropy of variable X is defined as the smaller number of bits needed to transmit a stream of symbols representing the values of X [16]. Information gain is calculated as $H(T) - H_S(T_i)$, where $H_S(T)$ is the weighted sum of the entropies for each subset T1, T2, … Tk of the training set partitioned by a specific node - attribute S. The candidate attribute that maximizes this measure is the best candidate for a split. Further splitting follows the same criterion.

The main strength of DT as compared to other algorithms is its low computational complexity while maintaining a very high level of accuracy. In

[17], DT-assisted TCP significantly increased bandwidth usage in wireless networks without deteriorating the good behavior of TCP over wired networks. In [14], this algorithm is used to dynamically select the most efficient TCP flavor (Cubic, BBR, and C2TCP) for the specific flow type and under specific network conditions.

### 3.1.2 Bayesian Classifier

Bayesian classifier is also of the computationally less complex but still very flexible and robust algorithm. It relies on probabilistic approach and provides classification based on prior knowledge and observed data. It does not only classify data, but it also gives probability distribution over all classes. The classifier can easily be updated with each training example.

Given the n attributes and j classes, the classification on the new data $(a_1, a_2, ... a_n)$ as belonging to the specific class $v_j$ ($v_j \in V$) relies on the assumption of conditionally independent attributes, i.e.,

$$P(a_1, a_2 ... a_n | v_j) = \prod_i P(a_i | v_j)$$

New data will belong to the class $v_c$, where:

$$v_c = argmax P(v_j) \prod_i P(a_i | v_j)$$

Primarily because of robustness, learning and classification speed, Bayesian classifier is a good candidate to be implemented as a patch to the existing CC algorithms. An implementation of Bayesian classifier given in [18] is a proof of concept showing the superiority of this approach over the comparable protocols. The advantages of Bayesian classifier over the popular SL methods such as Support Vector Machine, K Nearest Neighbor and Random Forest classifiers in CC for Heterogenous Vehicular Networks is shown in [19].

### 3.2 Remodeling CC - new dynamically adaptive CC schemes

The aforementioned SL algorithms use stored historical data to make a specific classification or prediction on new data. As shown, they can significantly improve the existing rule-based CC algorithms in many ways.

Another lane towards better CC might introduce a completely new network-specific and flow-specific CC paradigm. CC should not incorporate pre-defined universally applied rules. Instead, they should be adaptable to various networking environments and behave differently under different conditions.

RL has shown a great potential in dealing with highly dynamic priorly-unknown environments in real-time. In contrast to SL, RL algorithms can interact with the network and can update their CC actions in accordance with the received feedbacks in real-time. This makes this group of algorithms a good candidate for new real-time condition-adaptable CC models.

Agents explore a specific environment and learn by doing/moving, i.e., they learn which action is better to take under specific conditions in order to achieve a specific goal or in order to improve the system in accordance with a specific objective. In the context of CC, **state space** contains the CC-specific network parameters (such as delay, packet loss etc.), **action space** contains specific actions that may be taken to control the congestion (such as cwnd parameter variation), while the goals (such as achieving maximum throughput with minimum delays and packet loss) are incorporated into the **reward function**.

There are two main types of RL algorithms – **value-based** and **policy-based** RL. Each method has its advantages. Value-based RL algorithms are generally more sample efficient and show high efficiency in many environments, especially when they are used along with the deep networks. On the other hand, policy-based algorithms converge faster and are more suitable for continuous stochastic environments. However, high variance in the gradients leads to slower learning. The combination of two approaches has been recently studied extensively, and has proven to be superior over both fundamental approaches individually. In order to use advantages of both, **Actor-Critic (AC)** family of algorithms use both value-based and policy-based RL jointly. The system is composed of two main modules: an actor and a critic, which represent a policy and value-function, respectively. The actor takes a state as an input, and outputs a specific action. The critic "criticizes" the actor (i.e., it considers the specific state and evaluates the action by computing the value function). Being criticized, the actor updates its policy. Based on a taken action and on a received reward, the critic evaluates the prediction error and adjusts its own values. Consequently, both the actor and the critic learn from each other over time, which will improve the system output.

Based on the AC philosophy, a number of algorithms have evolved, and have been efficiently used to optimize the CC problem. In this mini-survey, we have selected just a few of them, based on their advantages and on the contribution to the CC remodeling.

### 3.2.1 Deep Q-Learning (DQL)

Q-Learning is a RL scheme that outputs the best action in accordance with an objective and under a specific state-condition. However, the complexity of Q-learning grows exponentially with the size of the state space, which makes the learning process time consuming and Q-table enormously large in complex environments. As such, straightforward implementation of Q-Learning in CC is unfeasible.

DQL deploys deep neural network (DNNs) to predict Q values, instead of learning them through iterations. By minimizing loss function $L(\theta_t)$ at time $t$ (through stochastic gradient descent), DNN with weight $\theta$ is used as function approximation for estimating Q-value function, i.e., $Q(s, a, \theta) \approx Q^*(s, a)$.

Where Q* is given by the Bellman equation:
$$Q^*(s, a) = E[r_{t+1} + \gamma \max \{Q^*_{a'}(s', a')\}]$$
where, $r$ denotes rewards, $\gamma$ is the discount factor

Deep learning assumes stationary data distribution and independent data samples. In order to maintain the stability and sampling efficiency, DQL needs to employ experience replay buffers and additional target network to estimate Q value, while online network predicts actions.

In the context of the DQL-based CC modeling, rewards are derived from a specific utility function. Utility function is based on the specific objectives. In CC the objectives can be defined as a function that increases with the increase of throughput and decrease of the delay. Then Q-value function is approximated using a specific deep learning algorithm. Finally, when a specific set of state variables (which reflect the network conditions at a given time) are passed to the currently trained policy, the algorithm will generate a cwnd-adjusting action from a defined action space.

The main advantage of using DQL in CC is its high efficiency in learning value functions and optimal policies in an on-line manner, with minimal computational complexity at the cost of somewhat slower convergence. Current successful implementations in IoT applications [20], Ad-Hoc Wireless networks [21], and in varying bandwidth networks [11] etc., show the aforementioned efficiency of DQL in CC.

### 3.2.2 Asynchronous Advantage Actor Critic Algorithm (A3C)

Actual AC-based implementations (DDPG, A3C, etc.) use neural network function approximators to design the actor and critic. The networks are trained for the actor to find its global maximum and for the critic to minimize its evaluation error over time. In contrast to the original policy gradient methodology (where the weights' update is made at the end of an episode), in AC, neural networks update their parameters at each step.

A number of updates have been applied to the original version of AC algorithm. Direct successors are Advantage AC (A2C) and Asynchronous AC (A3C). Both A2C and A3C represent Q function as a sum of estimate of the value function V (obtained from the critic network) and advantage function $A(s_t, a_t)$. Advantage function therefore represents the difference between the Q-value (of picking an action $a_t$ while being in state $s_t$) and the expected rewards for actions (derived from policy $\pi_\theta$) i.e.,

$$A(s_t, a_t) = r_t + \gamma V(s_{t+1}; \theta_v) - V(s_t; \theta_v)$$

where γ is the discount factor, and θ are parameters of the policy π.

Advantage function tells the agent how much an action turns out to be better or worse than expected, which helps the agent update its network more efficiently. Updating the policy gradient by using A(s,a) instead of Q(s,a) contributes to the reduction of high variance of the actor network.

The gradient of the cumulative reward is given with [22]:

$$\nabla_\theta E_{\pi_\theta} \left[ \sum_{t=0}^{\infty} \gamma^t r_t \right] = E_{\pi_\theta}[\nabla_\theta log \pi_\theta(s,a) A^{\pi_\theta}(s,a)]$$

A3C uses multiple copies of the same agent to independently explore the environment at the same time. They collect samples and compute policy gradients and value function gradients parallelly. The individually derived gradients are asynchronously passed to the global network, which then updates its actor and critic. From time to time, agents copy the weights of the global network to update their own weights. Working in parallel enables the exploration of the bigger state-action space for the same amount of time as compared to the previous AC version. Moreover, it eliminates the need for the Experience Reply Buffer.

Multi-agent structure contributes to the faster exploration of the state space, more accurate predictions, and faster convergence of the algorithm. We find this group of algorithms to be the method of choice in ML-based CC remodeling. A recent work that shows the efficiency of A3C as implemented in CC for different flow types and different networking environments is given in [23].

### 3.2.3 Deep Deterministic Policy Gradient (DDPG) and TD3

DDPG is an AC-based method designed to work well in continuous action spaces. Compared with DQL, DDPG has stronger capability to train models in more complex environments [1]. DDPG combines Q-learning and Policy gradients to output deterministic policy. Again, for better stability and learning efficiency, target network and circular experience buffer are used. In DDPG, both actor and critic are updated by two neural network structures. The actor uses one network to update the policy parameters and another one to select actions. The critic uses one network to update Q parameters and another to calculate Q values.

The critic loss is a MSE of time difference-error:
$$L(\theta^Q) = E[(Q_{target} - Q_{predict})^2],$$
$$L(\theta^Q) = E[r + \gamma Q(s', \mu(s', \mu(s'; \theta^\mu); \theta^Q) - Q(s, a; \theta^Q)]$$

The actor is updated using sampled policy gradient:
$$\nabla_{\theta^\mu} J^{\theta^\mu} \approx E[\nabla_{\theta^\mu} Q(s, \mu(s; \theta^\mu); \theta^\mu)] = E[\nabla_{\mu(s)} Q(s, \mu(s); \theta^Q) \nabla_{\theta^\mu} \mu(s; \theta^\mu)]$$

In the above equations, target networks are used to estimate Q-value for the next state while $Q_{predict}$ values use Q-network with newest weights. Target policy that maps states to a specific action is denoted μ(s), while $\theta^Q$ and $\theta^\mu$ are the weights of the critic (Q network) and actor deterministic policy network, respectively.

Original DDPG has been upgraded to many variants, while Twin Delayed DDPG (T3D) has shown a good potential in CC. T3D uses a more complex structure made of two critics and an actor, with two NNs per each, resulting in six neural networks overall. In this structure, T3D improves over the original DDGB by using Clipped double Q-learning, delayed policy and target updates, as well as target policy smoothing.

In the context of CC modeling, a representative example of using a variant T3D is presented in Orca [24]. The state space contains average values of various CC parameters, such as delays, loss, etc. Actions are products of current congestion window with $2^x$, where -2<x<2. Delivery rate, delay, and losses are incorporated into the reward function. The aforementioned T3D-based CC work is one of the most successful RL-CC implementations so far.

### 3.2.4 Proximal Policy Optimization (PPO)

PPO addresses the issues of the traditional policy gradient philosophy and improves some aspects of its derivative - Trust Region Policy Optimization (TRPO). To deal with learning step size issue of the traditional policy gradient methods, PPO limits the range of the objective function. Surrogate objective function replaces the log probability of the initial policy gradient formulation with the ratio between new and old policy, i.e.,

$$L(\theta) = E\left[\frac{\pi_\theta(s, a)}{\pi_{Old}(s, a)} \hat{A}^\pi(s, a)\right] = E[r(\theta) \hat{A}^\pi(s, a)]$$

TRPO further adds trust region constraint in order to limit the policy change, i.e., to assure better (a more trusted) local estimation of gradients, and to guarantee that the policy is continuously improving. In order to limit the trust region, TRPO uses average KL divergence, which makes it computationally complex because of the second order derivatives. Instead, most of the PPO implementations use the Clipped Surrogate Objective to integrate the stabilizing constraint into the objective function:

$$L^{CLIP} = E[\min(r(\theta)\hat{A}^{\pi}(s,a), clip(r(\theta), 1-\epsilon, 1+\epsilon)\hat{A}^{\pi}(s,a))]$$

where $\epsilon \in [0,1]$.

By limiting the $r(\theta)$, PPO gains approximately the same performance as TRPO but solves its problem of complexity. More details on the algorithm can be found in [25].

An efficient implementation of PPO in CC is given in [26] where the provided model outperformed other comparable algorithms in almost all categories (bandwidth variability, loss variability, link dynamics, etc.). The system has shown to be robust to environments outside the scope of its training.

## 4  A roadmap to the implementation of ML in CC

Traditional rule-based algorithms have problems with the efficiency and low bandwidth utilization, since they are not designed to act optimally under different network conditions. ML-based techniques presented in this paper show a great potential to overcome most of the issues of the traditional CC. However, the implementation of the ML-based algorithms is limited by various factors.

Competing algorithms that start their learning at different times, different environments, with different input-output space may suffer from fairness issue. Some algorithms may be greedier than others and can "learn tricks" to free up network capacity and leave little or no space for the others.

While there are other issues as well, one of the major limitations of the ML-based CC is computational complexity. Computational complexity leads to delayed reactions of the ML-based CC, which further influence the efficiency and even feasibility of the ML-based CC. Many ideas can be proposed, such as design of the lightweight models or reducing the computational complexity at the cost of accuracy, etc. But these approaches would provide a small upgrade in CC performances as compared to the existing algorithms. Hence, we propose two computing paradigms to solve the complexity issue, namely dataflow computing and GaAs technology.

## 4.1 Dataflow computing

The dataflow paradigm [27, 28] supplements the control-flow paradigm [29]. In control-flow, programs are written to micro-control the flow of data through the hardware, based on the von Neumann architecture. In dataflow, the programs configure the hardware based on FPGAs, and other mechanisms are used to move data through the hardware.

The dataflow paradigm could achieve speedups of 10, 100, or even 1000 times, compared to the control flow paradigm. At the same time, the power reduction could be about 20 times smaller. Precision could be improved significantly. Finally, the size of the chip gets reduced with a factor of up to 10 times.

Potentially, time-consuming loops with lots of data re-usability in each particular loop iteration, could benefit a lot from this paradigm. This property has recently been found very useful for deep learning. Hence, many machine learning problems are moving towards the dataflow computing implementation space. Some examples of linear regression, decision trees and other algorithms can be found in appgalery.maxeler.com and [26]. We expect dataflow computing to become an important paradigm in practical ML-based CC implementations.

## 4.2 GaAs

The GaAs technology is an option for efficient design of processors and implementation of time-consuming algorithms. It offers a considerably higher computing speed and a much higher level or radiation hardness, so it could be used in end-to-end congestion control. On the other hand, the amount of logic that could be placed on a single chip is smaller, while the gate delay depends heavily on the fan-out of the gate.

The enumerated characteristics define the specific requirements of processor architecture and algorithm implementation. Also, the ratio of off-chip to on-chip delays is relatively high. This requires the utilization of highly pipelined architectures in which pipeline stages are of a relatively small complexity.

Relevant concepts on GaAs technology were described in [30] and [31]. Some efforts on GaAs processors, under DARPA sponsorship, were described in [32] and [33]. It follows that most effective implementations could be expected from algorithms that could be implemented using many small elements, connected in a pipelined fashion, such as perceptron, SVM, and similar approaches. We consider that ML-based CC algorithms can also benefit from the implementation of this technology.

## 5 Conclusions

This mini-survey reviews the deployment of the representative machine learning algorithms based on their potential to be used in congestion control and their up-to-date contribution to this field. In the selection process, the frequency of usage of each ML technique in solving the CC issues were also taken into account. The advantages of each of the algorithms as used in the context of CC are highlighted, and proof of concept is given for each of them.

The analysis also brings up many issues related to the implementation of ML in CC. Among them, computational overhead, energy consumption, and response time might be the hardest to overcome.

Instead of decreasing the algorithms' complexity at the cost of accuracy, we propose a different course to solve the complexity issue. Precisely, based on the structural and functional features of dataflow computing and GaAs technology, we envision that deep learning-based CC may greatly benefit from these paradigms.

Finally, this survey opens up new research discussions on the synergy square: communications-algorithms-architectures-technologies.